\documentclass[%
mathleft,%
final,%
]{an}
\usepackage{graphicx}
\usepackage{times}
\begin{document}

% The following seven commands are intended for editorial usage and
% should be ignored by the author(s).
\Pagespan{1}{}% Document's page range.
% If second parameter is left empty, the last page is computed
% automatically.
\Yearpublication{2013}%
\Yearsubmission{2012}%
\Month{1}%
\Volume{334}%
\Issue{1}%
\DOI{This.is/not.aDOI}%

\title{Disks, accretion and outflows of brown dwarfs}
%Cool Stars 17 proceedings in Astronomische Nachrichten -- \\
%        template file for authors using \LaTeXe\ \thanks{Data
%were obtained with the STELLA facility}}

\author{V. Joergens\inst{1,2}
%\fnmsep
%\thanks{Corresponding author. \email{viki@mpia.de}}
% Example for footnote, note the usage of the \fnmsep command
% as separator between institute number and footnote mark}
\and G. Herczeg\inst{3}
\and Y. Liu\inst{4}
\and I. Pascucci\inst{5}
\and E. Whelan \inst{6}
\and J. Alcal\'a \inst{7}
\and K. Biazzo \inst{7}
\and G. Costigan \inst{8}
\and M. Gully-Santiago \inst{9}
\and Th. Henning \inst{2}
\and A. Natta\inst{10}
\and E. Rigliaco \inst{5}
\and V. Rodriguez-Ledesma \inst{2}
\and A. Sicilia-Aguilar\inst{11}
\and J. Tottle \inst{12}
\and S. Wolf \inst{4}
}
\titlerunning{Disks, accretion and outflows of brown dwarfs}
\authorrunning{V. Joergens et al.}
\institute{
Zentrum f\"ur Astronomie Heidelberg, Inst. f\"ur Theor. Astrophysik,
Albert-Ueberle-Str. 2, 69120 Heidelberg, Germany
\and
Max-Planck Institut f\"ur Astronomie, 
K\"onigstuhl~17, 69117 Heidelberg, Germany, viki@mpia.de
\and
%Greg:
Kavli Institute for Astronomy and Astrophysics, Peking University, Beijing, 100871, PR China
\and
%Yao:
University of Kiel, Institute of Theoretical Physics and Astrophysics, Leibnizstrasse 15, 24098 Kiel, Germany
\and
%Ilaria:
Lunar \& Planetary Lab.,
Depart. of Planetary Sciences,
Univ. of Arizona, 1629 E. University Blvd., Tucson, AZ 85721
\and
%Emma:
Dublin Institute for Advanced Studies,
10 Burlington Road, Dublin 4, Ireland
\and
%Juan:
INAF - Osservatorio Astronomico di Capodimonte, via Moiariello 16, 80131, Napoli, Italy
\and
%Grainne Costigan:
European Southern Observatory, Karl-Schwarzschild-Str. 2, 85748 Garching, Germany 
\and
%Michael:
Department of Astronomy, University of Texas at Austin, 1 University Station 
C1400, Austin, TX, 78712, USA 
\and
%Antonella:
Osservatorio Astrofisico di Arcetri, INAF, Largo E. Fermi 5, 50125 Firenze, Italy
\and
%Aurora:
Departamento de F{\'i}sica Te{\'o}rica,
Facultad de Ciencias, Univ. Aut{\'o}noma de Madrid,
Cantoblanco, 28049 Madrid, Spain
\and
%Jonathan: 
Imperial College London, 1010 Blackett Lab., Prince Consort Road, London SW7 2AZ, UK
}

\received{XXXX}
\accepted{XXXX}
\publonline{XXXX}


\abstract{Characterization of the properties of young brown dwarfs are important to constraining the formation of objects at the extreme low-mass end of the initial mass function.  While young brown dwarfs share many properties with solar-mass T Tauri stars, differences may be used as tests of how the physics of accretion/outflow and disk chemistry/dissipation depend on the mass of the central object.  
This article summarizes the presentations and discussions during the 
splinter session on 
{\em Disks, accretion and outflows of brown dwarfs} held at the CoolStars17 conference  
in Barcelona in June 2012.
Recent results in the field of brown dwarf disks and outflows
include the determination of brown dwarf disk masses and geometries based on 
Herschel far-IR photometry (70-160\,$\mu$m), accretion properties
based on X-Shooter spectra, and new outflow detections in the very low-mass regime.
}

\maketitle

\section{Introduction}
\label{sect:intro} 

The exploration of disks, accretion and outflows of young brown dwarfs (BD)
plays an important role in developing our understanding of BD formation,
planet formation, and the physics of circumstellar disks and outflows in general.
It is related to fundamental open questions in stellar astronomy, such as: Do BDs form via the same path as stars?
Can planets form around BDs? How do disks develop in a low-gravity, -temperature, and -radiation environment?.

It has been established in the last years that 
BDs and very low-mass stars (VLMS) at an age of a few Myrs
resemble higher mass T~Tauri stars. They exhibit 
surface activity, such as cool spots (e.g., Joergens et al. 2003; Rodr\'iguez-Ledesma et al. 2009)
and coronal activity
(e.g., Stelzer et al. 2006). BD/VLMS have disks with a similar fraction as stars detected
at mid-IR (e.g., Comer\'on et al. 2000; Natta \& Testi 2001;
Jayawardhana et al. 2003; Luhman et al. 2008) and
far-IR/submm wavelengths (e.g., Klein et al. 2003; Scholz et al. 2006; Harvey et al. 2012),  
which are actively accreting (e.g., White \& Basri 2003;  Mohanty et al. 2005; Herczeg \& Hillenbrand 2008; 
Rigliaco et al. 2011), and often show signs of grain growth and crystallization (e.g., Apai et al. 2005; 
Pascucci et al. 2009).
The finding that very young BDs rotate on average slower 
(e.g., Joergens \& Guenther 2001; Joergens et al. 2003; Caballero et al. 2004) 
than their older counterparts (e.g., Bailer-Jones \& Mundt 2001; Mohanty \& Basri 2003) 
is indicative of a magnetic braking mechanism due to interaction with the disk.  Outflows from BD/VLMS 
have been observed in
spectro-astrometry of forbidden emission lines (e.g., Whelan et al. 2005, 2009;
Bacciotti et al. 2011; Joergens et al. 2012a), 
and in images of CO J=2-1 emission (Phan-Bao et al. 2008).

Many details of the properties of BD disks and outflows, however, are still unclear:
What are the masses and sizes of BD/VLMS disks?
How does grain evolution occur in these disks? Do planets form in circum-BD disks?
What are the properties of BD/VLMS outflows and how does the outflow mechanism work
at low mass-accretion rates? Are there any discontinuities of all these properties
with mass indicating a different formation path of BDs compared to stars? 
Several new instruments that are ideally suited to study 
disks and outflows of faint objects (PACS/Herschel, X-Shooter/VLT, ALMA-early science)
are producing currently their first results.
At the splinter session on
{\em Disks, accretion and outflows of brown dwarfs} at the CoolStars17 conference
(Barcelona, June 2012, www.mpia.de/homes/joergens/cs17splinter.html)
leading scientists as well as young researchers 
presented and discussed recent results in the field of BD/VLMS disks and outflows, as outlined in the 
following.

\section{Accretion in BD disks}
\label{sect:acc}

Stars and BDs are thought to gain most of their mass through disk 
accretion. This process also regulates 
the angular momentum that the object receives from the 
surrounding interstellar medium. 
In addition, accretion measurements provide a crude probe of 
the gas content in the disk. 
In the current picture of magnetospheric accretion, 
material accreted through the disk is falling along magnetic accretion columns 
onto the stellar surface causing an accretion shock 
(e.g., Ghosh \& Lamb 1979; K\"onigl 1991; Shu et al. 1993).
For solar mass stars, the evidence
for magnetospheric accretion in funnel flows is based primarily on
long-term monitoring of line profiles of individual stars (Bouvier
et al.~2007; Alencar et al. 2012), models of these line profiles
(Kurosawa et al. 2008; Fischer et al. 2008), and models of how
accretion would flow in magnetospheric structures constructed from
Zeeman Doppler imaging (e.g., Gregory \& Donati 2011).  The
magnetospheric accretion geometry is assumed to be similar for BDs
as for classical T~Tauri stars (CTTS), with some support from models of H$\alpha$
line profiles (Muzerolle et al. 2003, 2005).

Accretion rates, derived from the accretion luminosity,
\begin{equation}
\dot{M} = \left( 1 - \frac{R_*}{R_{in}} \right ) ^{-1} \frac{L_{acc} R_*}{G M_*} \sim 1.25 ~\frac{L_{acc} R_*}{G M_*}
\end{equation}
can be measured directly from the excess continuum emission, which is
predominantly emitted from the shock regions,
or indirectly from a variety of tracers, 
e.g. hydrogen and helium recombination lines, which are
produced predominantly in the accretion funnel flows (Gullbring et al. 1998).  
VLT/X-Shooter observations covering the 
300-2500\,nm wavelength range demonstrate the reliability in using many different 
indirect measures of accretion luminosity, including H I, He I, Na I, and Ca II lines, 
and with a similar ratio of line to continuum luminosity across the mass range from BDs to stars 
(Rigliaco et al.~2012; Fig.\,1). %\ref{fig:acc}).

Mass accretion rates for BDs are of the order of 
$10^{-9}$ to $10^{-12}\,M_{\odot}\,\mathrm{yr}^{-1}$
(e.g., Muzerolle et al. 2003; Natta et al. 2004; Mohanty et al. 2005; Herczeg et al. 2009).  The accretion rate scales with the mass of the central object
from stars of several solar masses to 0.1\,$M_{\odot}$ 
as $\dot{M} \propto M^{\alpha}$ with $\alpha$=1.3-3  
(Muzerolle et al. 2005; Natta et al. 2006; Garcia-Lopez et al. 2006; Fang et al. 2009; Biazzo et al. 2012, Manara et al.~2012).
Recent multi-epoch spectroscopy of BD/VLMS in Cha\,I revealed a population of Class~II 
sources with no significant accretion (Mohanty \& Tottle 2013, in prep.).  They can be separated into 3 classes:
(1) edge-on disks, which artificially suppresses the  
H$\alpha$ emission by disk occlusion of the accretion funnel flows/hot spots; 
(2) disks with large inner holes, in which accretion is truly quenched; and 
(3) flat anemic disks, in which the reduction of both the disk flux and accretion rate is presumably 
simply a result of evolution.

% ---------------- Fig 3
\begin{figure}
\includegraphics[width=7cm, clip]{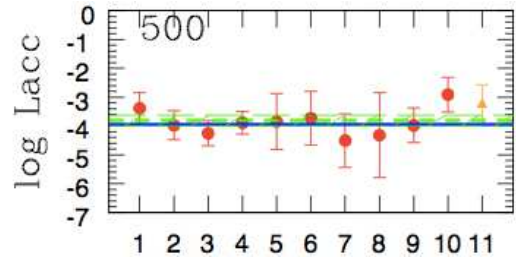}
\label{fig:acc}
\caption{
Accretion indicators based on X-Shooter observations
from Rigliaco et al. (2012). Displayed are the accretion luminosity Lacc
derived for a BD based on 11 different accretion indicators
(1=H$\alpha$, 2-H$\beta$, 3=H$\gamma$, 4=H9, 5=Pa$\gamma$, 6=Pa$\beta$, 7=HeI, 8=NaI, 9=Ca854, 10=866, 11=Uband);
green dashed line: average Lacc computed considering the accretion tracers, blue solid line: Lacc from the
excess continuum emission.
}
\label{fig:xshooter}
\end{figure}

Variability is an essential %but poorly constrained 
part of the accretion process for T~Tauri
objects from solar to substellar masses (e.g., Stelzer et al. 2007). 
Study of its origin is important to understand the underlying physics 
and correctly interpret the observed signatures of e.g. accretion.
Recent spectroscopic monitoring with FLAMES/VLT of 25 stars (G2-M5.75) in Cha\,I 
covering time-scales of weeks and months revealed variable 
accretion rates with average amplitudes of 0.89\,dex based on Ca\,II and 0.41\,dex 
based on H$\alpha$ EW, resp. (Costigan et al. 2012).
A significant portion of this accretion variability occurs on short timescales (8-25\,days),
which are of the order of the rotation rates of such objects 
(e.g., Joergens et al. 2003; Rodriguez-Ledesma et al. 2009), 
indicating that it probes the spatial structure in the accretion flows.
In rare cases, large amplitude photometric variability in BDs is detected and attributed 
to periodic obscuration of the central object by the circum-BD or circumbinary disk (Looper et al. 2010; 
Cody \& Hillenbrand 2011; Rodriguez-Ledesma et al.~2012).
For the BD candidate CHS7797 (M6), this is shown impressively based on  
I-band monitoring covering 6\,yrs as well as
simultaneous monitoring in the I, J, Ks, IRAC [3.6], and [4.5] 
bands over 40\,days (Rodriguez-Ledesma et al.~2012).

\section{Disk masses and geometries of BDs}

The existence of dust disks around sub-stellar objects is a well-established 
phenomenon, but is mainly based on the detection of mid-IR excess emission (cf. Sect.\ref{sect:intro}),
which can probe only about the inner $\sim$0.1\,AU around BDs.
The high sensitivity of the \emph{Herschel Space Observatory} and its 
coverage of the far-IR portion of the wavelength regime (70, 160\,$\mu$m)
allows for the first time to trace cold dust (radial extents of $\sim$1\,AU) and 
constrain the disk properties for a statistically significant sample of young BDs. 
A Herschel survey of 47 BD/VLMS (M3-M9.5) in young star-forming regions, 
for which there is evidence of circumstellar dust from Spitzer (8-24\,$\mu$m),
detected 77\% of them at 70\,$\mu$m and 30\% at 160\,$\mu$m (Harvey et al. 2012).
The disk properties that were derived based on modeling of their spectral energy distribution 
using radiative transfer codes (Wolf 2003; Pinte et al. 2009) showed that
these BD/VLMS have very low disk masses (10$^{-6}$-10$^{-3}\,M_\odot$ = 0.3\,M$_{\oplus}$-1\,M$_{Jup}$, 
Fig.\,2) %\ref{fig:diskmass})
compared to disks of CTTS (10$^{-3}-10^{-1}\,M_\odot$, e.g. Williams \& Cieza 2011). 
Such low disk masses were also derived for other BD/VLMS based on far-IR/submm data
(e.g., Joergens et al. 2012a).
While lower disk masses are expected for objects of lower mass, 
the Herschel survey hints that the ratio M$_{disk}$/M could be smaller 
for BD/VLMS by a factor of 3-10 compared to that of 
CTTS and that they cover a wider range. 
The geometries (scale heights, flaring angles) of BD/VLMS disks are found to be consistent 
with those seen around CTTS,
although they are not as flared as expected based on gas being in hydrostatic equilibrium (Sz{\H u}cs et al. 2010).

The disk masses from Harvey et al. (2012) may be affected by a potential bias towards weak BD 
disks in the Herschel sample because of selection against targets in regions covered by  
Herschel's large scale shallow surveys.
The dust masses also rely on the standard assumption of a gas-to-dust ratio of 100.
While Harvey et al. (2012) show that
at $\geq$160\,$\mu$m, the disks are typically optically thin over most of the radii emitting 
at those wavelengths, 
ALMA observations will trace colder dust than observed with Herschel and may also provide direct 
measurements of the gas content in disks and will, therefore, be important to determine 
both the masses and sizes of BD disks.

% ---------------- Fig 2
\begin{figure}
\includegraphics[width=6.2cm, clip]{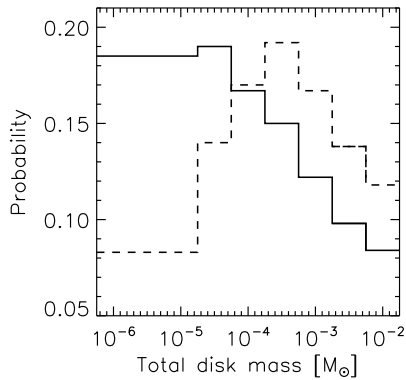}
\label{fig:diskmass}
\caption{Distribution of BD/VLMS disk masses determined based on SED fitting including Herschel data 
(from Harvey et al. 2012).}
\end{figure}

\section{Evolution of gas and grains in BD disks}

There is a growing observational evidence that disks around stars of different 
masses evolve differently. Dust disk lifetimes, as measured from the presence of IR excess 
emission, appear to be longer toward lower mass stars and BDs (Carpenter et al. 2006; Luhman 2009). 
These comparisons rely on identification of diskless BDs, including those found in ongoing surveys 
that target water absorption bands in BD atmospheres (Gully-Santiago et al. 2013, in prep).  
The processes of grain growth and dust settling, which are the first steps toward 
planet formation, appear, on the contrary, to occur faster in disks around BD/VLMS (Apai et al. 2005; 
Allers et al. 2006; Pascucci et al. 2009; Riaz 2009; Sz{\H u}cs et al. 2010; Riaz et al. 2012).  
This is based on the finding of
differences in the scale-height of the gas (from the temperature distribution) and 
that of the grains (from the SED) between CTTS and BD.
While the evolution of the gas is not constrained, the steep dependence between mass accretion rate and 
mass of the central star (cf. Sect.\,\ref{sect:acc}) could be the sign of a faster disk clearing around BD/VLMS. 
In the picture proposed by Hartmann et al. (2006) these disks are fully magnetically active 
(unlike disks around sun-like stars that have dead zones), hence they accrete faster. Finally, we 
discussed the chemistry occurring on the surface of BD disks, with emphasis on the low abundance of 
HCN and the depletion of water with respect to disks around sun-like stars 
(Pascucci et al. 2009, Pascucci et al. 2013, in prep).

\section{Outflows of BDs}

% ---------------- Fig 
\begin{figure}
\includegraphics[width=6.1cm,clip]{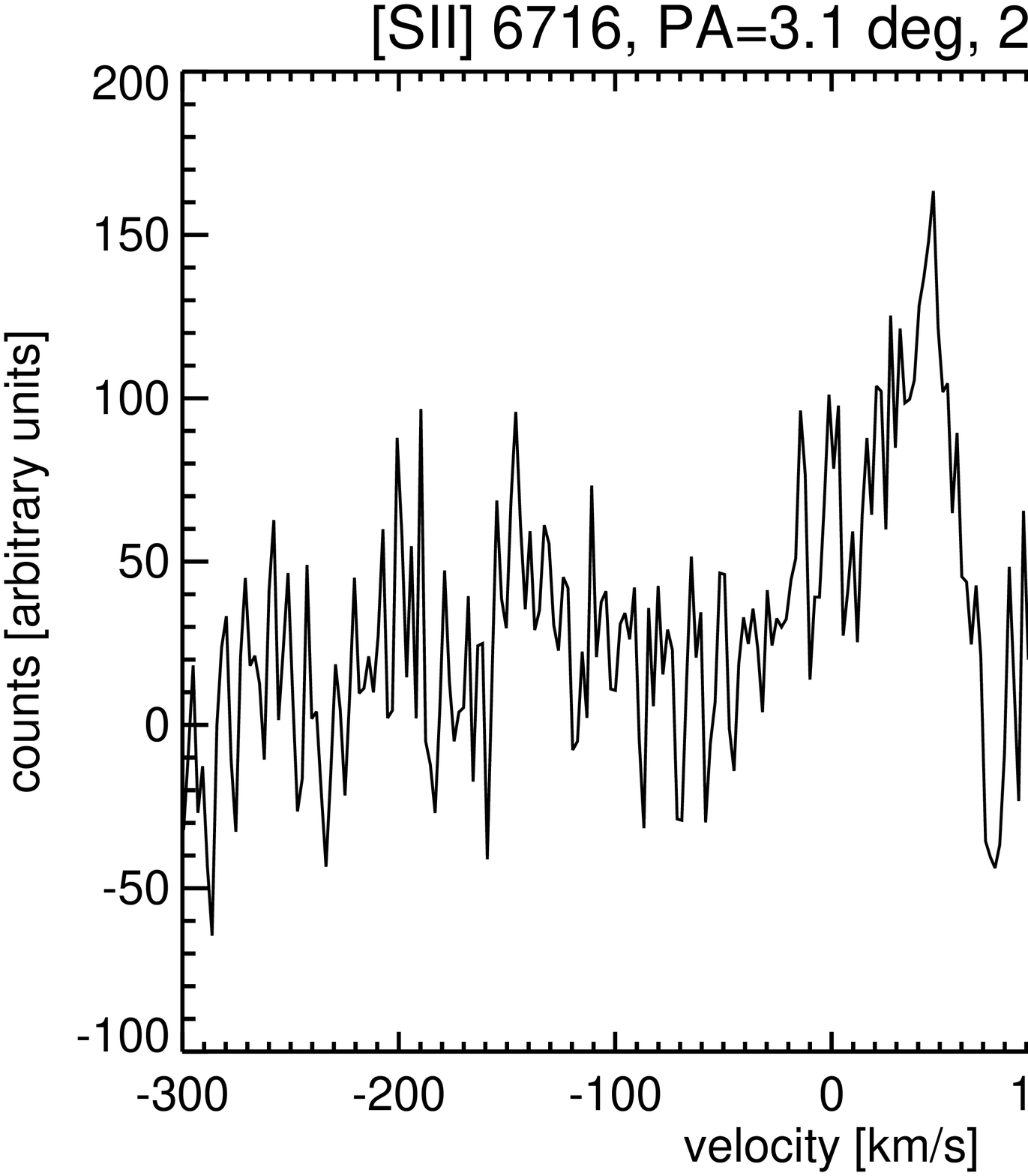}
\includegraphics[width=6.1cm,clip]{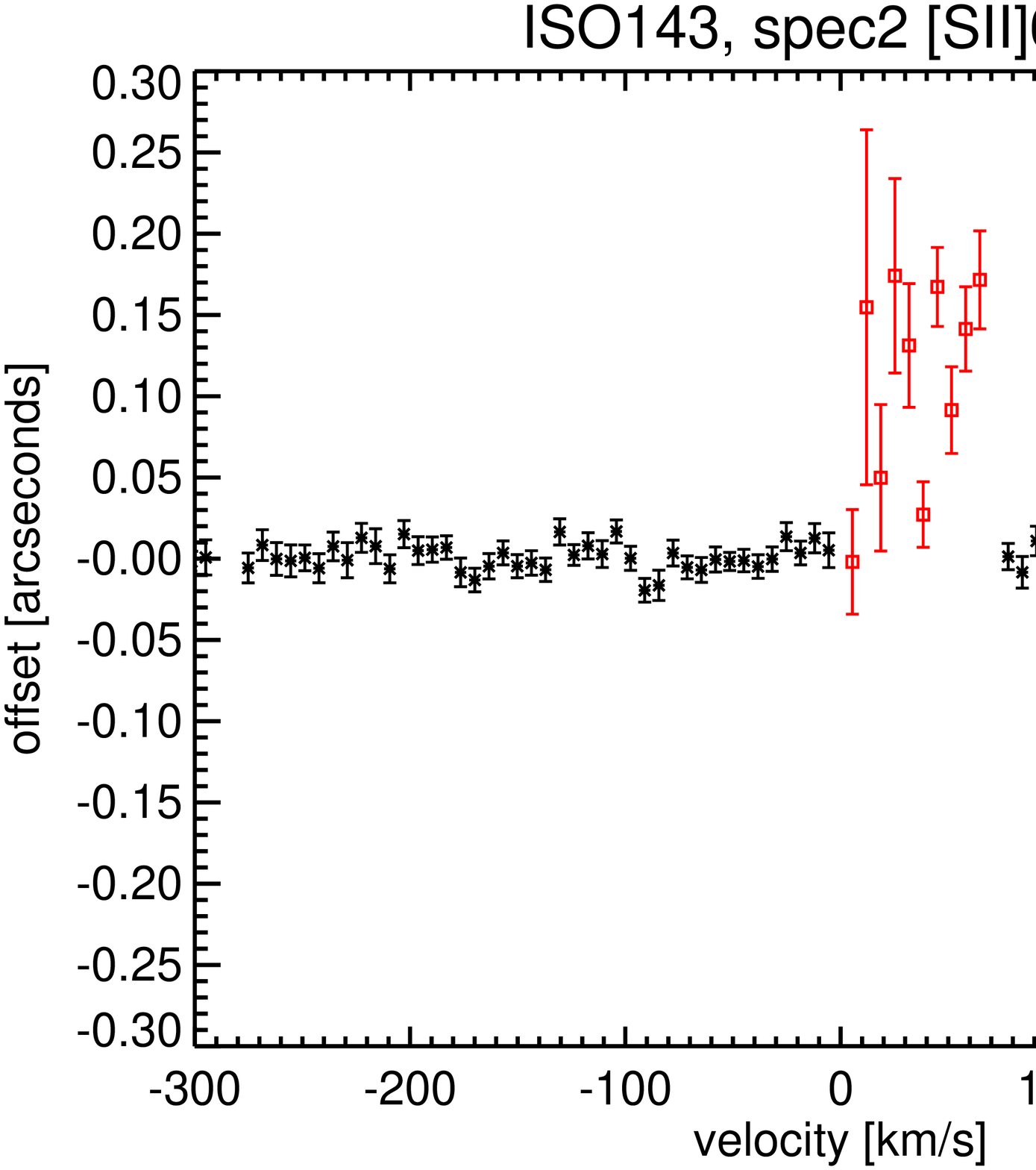}
\label{fig:outflow}
\caption{Discovery of an asymmetric outflow of a VLMS star (M5) by 
spectro-astrometry of forbidden emission of [S\,II]\,$\lambda$6716
(from Joergens et al. 2012b).
\textit{Top}: 
Line profile. \textit{Bottom}: Spatial offset vs. radial velocity
of the continuum (black asterisks) and of the continuum subtracted FEL (red squares)
demonstrating that the [S\,II] line is emitted at a distance of up to 170\,mas ($\sim$30\,AU) from the central source 
with velocities of up to 65\,km\,s$^{-1}$.
}
\end{figure}

The launch and driving of jets and outflows is a key process in the formation phase.
Outflows carry away both mass and angular momentum from the star-disk system 
(e.g., Launhardt et al. 2009).
In recent years, outflows driven by objects as low-mass as BD/VLMS were
found; the majority of these are optical outflows discovered by
spectro-astrometry of forbidden emission lines (FELs) formed in shocks 
(e.g.,  Whelan et al. 2005, 2009; Bacciotti et al. 2011; Joergens et al. 2012a), 
alongside a spatially resolved molecular outflow detected in the CO J=2-1 transition (Phan-Bao et al. 2008, 2011).
While this research is 
in an early stage and the number of BD/VLMS outflows confirmed by spatial information 
(spectro-astrometry or direct resolution)
is still
statistically small ($<$10), comparison between CTTS and BD/VLMS 
outflows has yielded a number of interesting results (Whelan et al. 2009; 
Bacciotti et al. 2011; Joergens et al. 2012a).  
Similarities between outflows of CTTS and BDs include, the observation of
asymmetries (Joergens et al. 2012b; Fig.\,3), %\ref{fig:outflow}), 
both low- and high-velocity
emission (Whelan et al. 2009), and jet collimation  (Whelan et al. 2012).  Initial
measurements of the ratio between the mass outflow and accretion rates $\dot{M}_{out}$/$\dot{M}_{acc}$
in BD/VLMS (Comer\'on et al. 2003; Whelan et al. 2009; Bacciotti et al. 2011)
hint that it might be higher than the 5--10\% in CTTS
(e.g., Hartigan et al.~1995; Sicilia-Aguilar et al. 2010; Fang et al. 2009).
If this is confirmed, it would point to different efficiencies in the mechanisms 
at work to launch outflows at very low masses or very low mass accretion rates. 
The molecular outflow detected by Phan-Bao et al. (2008) is around an
optically visible VLMS, i.e. an assumed counterpart to a Class~II CTTS, while 
molecular outflows in the stellar regime are usually associated with Class~0 and Class~I
objects.
The colder environment of BD/VLMS may mean that more
molecular material is available in the later stages of evolution and, therefore,
that CO outflows will be more commonly observed. 
The work on BD/VLMS outflows is now at a
stage where techniques for detecting them are well developed and
understood. First results have brought up some interesting questions, which 
need to be addressed through detections and study of BD/VLMS outflows for a larger
number of objects.

\acknowledgements
VJ acknowledges funding by the ESF in Ba.-W\"u.
YL acknowledges funding by the German Academic 
Exchange Service and the Research Unit FOR 795.

% Use this code if you wish to generate your bibliography with BibTeX;
% please replace first the string "an-demo" below with the name(s) of
% the BibTeX data base(s) you want to use.
% The resulting bibliography-output (the contents of the .bbl file)
% must be pasted into this file before submission.
%
% \bibliographystyle{an}
% \bibliography{an-demo}
%
% Replace the following example bibliography with your references
% before submission:

\end{document}